\documentclass{aastex62}

\usepackage{amsmath}

\received{}
\revised{}
\accepted{}

\shorttitle{Dissipation scale lengths of solar wind turbulence}
\shortauthors{Sasikumar Raja et al.}

\begin{document}

\title{Dissipation scale lengths of density turbulence in the inner solar wind}

\correspondingauthor{K. Sasikumar Raja}
\email{sasikumar@prl.res.in, sasikumarraja@gmail.com}

\author[0000-0002-1192-1804]{K. Sasikumar Raja}
\affil{Physical Research Laboratory, Navrangpura, Ahmedabad-380 009, India.}
\altaffiliation{Formerly at Indian Institute of Science Education and Research, Pashan, Pune - 411 008, India}

\author{Prasad Subramanian}
\affiliation{Indian Institute of Science Education and Research, Pashan, Pune - 411 008, India}
\author{Madhusudan Ingale}
\affiliation{Indian Institute of Science Education and Research, Pashan, Pune - 411 008, India}

\author{R. Ramesh}
\affiliation{Indian Institute of Astrophysics, 2nd Block, Koramangala, Bangalore - 560 034, India.}




\begin{abstract}

Knowing the lengthscales at which turbulent fluctuations dissipate is key to understanding the nature of weakly compressible magnetohydrodynamic turbulence. We use radio wavelength interferometric imaging observations which measure the extent to which distant cosmic sources observed against the inner solar wind are scatter-broadened. We interpret these observations to determine that the dissipation scales of solar wind density turbulence at heliocentric distances of 2.5 -- 20.27 $R_{\odot}$ range from $\approx$  13500 to 520 m. Our estimates from $\approx$ 10--20 $R_{\odot}$ suggest that the dissipation scale corresponds to the proton gyroradius. They are relevant to {\it in-situ} observations to be made by the Parker Solar Probe, and are expected to enhance our understanding of solar wind acceleration.

\end{abstract}



\section{Introduction} \label{sec:intro}

The solar wind has been long recognized as an excellent laboratory for studying the phenomenon of turbulence in magnetized plasmas. Among the enduring questions in turbulence studies is the mechanism by which fluctuations dissipate at small scales. Apart from basic interest in this issue, it enables us to estimate proton heating and viscous dissipation, which influence the aerodynamic drag experienced by solar coronal mass ejections and the acceleration of the solar wind. A knowledge of dissipation processes at these distances can impact our understanding of the manner in which the solar wind is accelerated. The route to understanding dissipation mechanisms starts with an identification of the lengthscales at which the (largely) self-similar turbulent spectrum steepens. There has been considerable recent interest in this area, both by way of observations near the Earth \citep{Saf2015, Che2012, Ale2008, Mal2010} and by way of hybrid fluid-kinetic simulations \citep{Fra2015, Tol2015, Cer2016, Cer2017}. We approach this issue using imaging observations of the angular broadening of distant cosmic sources observed (at radio frequencies) against the turbulent solar wind. Density fluctuations in the turbulent solar wind result in refractive index fluctuations which cause the background object appearing larger and ``fuzzier'', much in the way a car's headlights appear when viewed through fog. The images are constructed using interferometric correlations, (called visibilities), which can be related to the concept of a structure function. The structure function 
quantifies the two point correlation of phase corrugations introduced by the turbulent density inhomogeneities in the solar wind. In turn, it can be related to the spatial power spectrum of the turbulent density inhomogeneities. This connection enables us to constrain the spatial scale at which the spatial power spectrum starts to steepen, which is the dissipation scale. Our treatment takes into account possible anisotropies in the scattering process, and does not appeal to any specific physical models for the dissipation scale. While many observational attempts at quantifying the dissipation scale have used measurements near the Earth (which is $\approx$ 215 $R_{\odot}$ from the Sun), our results are among the few obtained for the inner solar wind, at distances ranging from 2.5 to 20.27 $R_{\odot}$ from the Sun.

\section{Observations}\label{observations}

\begin{figure}[!ht]
\centering
\includegraphics[scale=0.45]{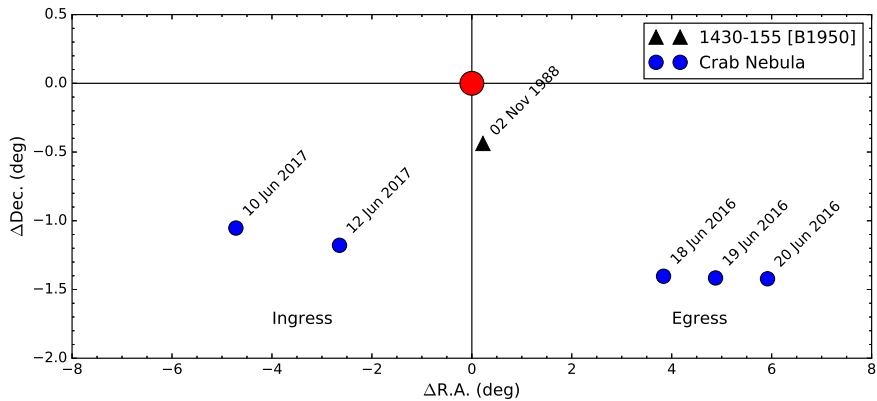}
\caption{
The red circle indicates the position of the Sun. The relative 
offsets of the Crab Nebula and 1430-155 from the Sun 
are marked by the blue circles and a black triangle respectively.
}
\label{fig:schematic}
\end{figure}

\begin{figure}[!ht]
\centering
\includegraphics[scale=0.45]{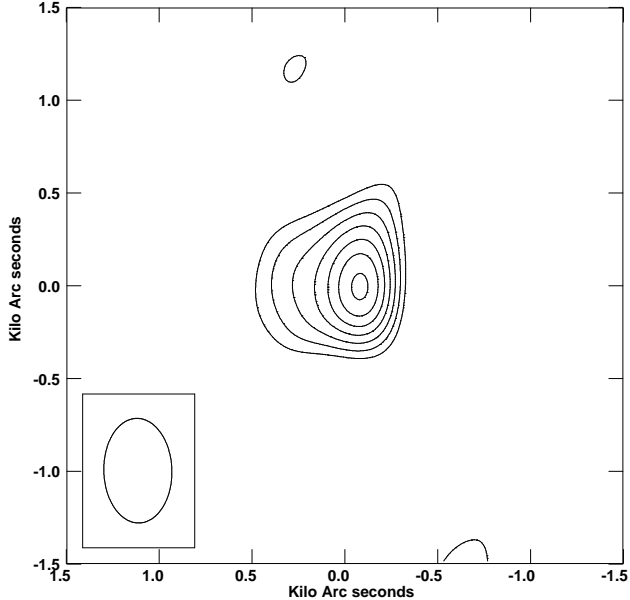}
\caption{The scatter-broadened image of the Crab Nebula observed using GRAPH on 18 June 2016. 
    It was observed at 80 MHz and the projected heliocentric distance was $\approx 13.46~R_{\odot}$. The observed flux density was $\approx 1293$ Jy with an axial anisotropy $\approx 1.4$.
The contour levels shown are 3, 4, 5, 6, 7, 8, 9
times of the rms noise ($\approx 140$ Jy). The convolving beam of the GRAPH 
is shown in the 
box located at lower-left of the image. 
}
\label{fig:aips_image}
\end{figure}

Our observations involve sources which were unresolved when observed far from the Sun (for the instruments with which we observed them).
However, when these sources are observed through the turbulent screen of the solar wind relatively close to the Sun, they become scatter-broadened; e.g., \citep{Sas2016, Sas2017, Ram2001}, and the extent of this broadening can help us constrain the dissipation scale-lengths of the turbulent spectrum. We use 3.65 cm observations of the source 1430-155 (B1950) made with the Karl G. Jansky Very Large Array (VLA) \citep{Tho1980, Per2011} and observations of the Crab nebula at 80 MHz using the upgraded Gauribidanur RAdio heliograPH (GRAPH) \citep{Ram1998, Ram2011}. Figure \ref{fig:schematic} shows the projected heliocentric distances of each of the radio sources we have used in our analysis. Figure \ref{fig:aips_image} shows a representative scatter-broadened image of the Crab Nebula observed at a heliocentric distance of  $\approx 13.46~R_{\odot}$ on 18 June 2016. The image is markedly anisotropic; the ratio of the major to minor axis ($\rho$) is 1.4 and the position angle of the major axis (PA, measured from north through east) is $\approx 117^{\circ}$. Details of all the observations are given in Table~\ref{tab:one}.

\section{Results and Discussions}

\subsection{Estimating the dissipation scale}\label{sec:sas1}

The building block for obtaining radio images is the quantity $\Gamma(s)$ which measures the spatial coherence of the electric field detected by a pair of antennas separated by a distance $s$: 

\begin{equation}\label{mf}
\Gamma(s)={V(s) \over V(0)} = {\langle E(r)E^*(r+s) \rangle \over \langle |E|^2 \rangle}
\end{equation}
The quantity $V(s)$, called the visibility, is the time-averaged correlation between electric fields detected by a pair of antennas separated by a distance $s$ and $V(0)$ is the so-called ``zero-spacing'' visibility. Each pair of antennas in an interferometric array yields a visibility, which (by the van Citterte-Zernicke theorem) is a point in the (spatial) Fourier transform of the scatter-broadened image of the object. The image (such as the one shown in figure~\ref{fig:aips_image}) is constructed by suitably inverse Fourier transforming a set of calibrated visibilities using the Astronomical Image Processing System (AIPS)\footnote{http://www.aips.nrao.edu/index.shtml}. The structure function $D_{\rm obs}$ is related to $\Gamma (s)$ via 
\begin{equation}
D_{\rm obs}(s)=-2 \ln \bigg [ \frac{V(s)}{V(0)} \bigg ] \, ,
\label{obs_structfun}
\end{equation}
where the subscript ``obs'' serves as a reminder that it is derived from the calibrated visibilities that form the observed image.
On the other hand, the theoretical view of the structure function relates it to the spectrum of turbulent fluctuations that give rise to the observed scatter broadening. The spectrum of turbulent density fluctuations ($P_{\delta n}$) in the solar wind is commonly modelled as a power law with an exponential falloff at the inner/dissipation scale \citep{Bas1994, Bas1995, Arm1995, Ale2012, Ing2014, Sas2016, Sas2017}. We account for anisotropy in the scatter-broadening process by allowing for different wavenumbers along ($k_{x}$) and perpendicular ($k_{y}$) to the large scale magnetic field \citep{Ing2014}:
\begin{eqnarray}
P_{\delta n}(k, R) = C_{N}^{2}(R) (\rho^2 ~k_x^2+k_y^2)^{-\alpha/2} \\ \nonumber
 \times ~ exp\bigg\{-(\rho^2 ~k_x^2+k_y^2)\bigg({l_{i}(R) \over 2 \pi }\bigg)^{2}\bigg\} \,  ,
\label{eq1}
\end{eqnarray}
where $R$ is the heliocentric distance, $C_{N}^{2}$ represents the amplitude of the density spectrum and $\rho \equiv k_{x}/k_{y}$ is interpreted as the ratio of the major and minor axes in the scatter-broadened image. The inner scale is also assumed to be anisotropic, with $l_{ix} = \rho l_{iy}$ and $l_{i} \equiv \sqrt{l_{ix}^{2} + l_{iy}^{2}}$. The slope ($\alpha$) of the turbulent power spectrum is well known to follow the Kolmogorov scaling law ($\alpha = 11/3$) at large scales (small wavenumbers). It is known to flatten near the dissipation scale ($\alpha = 3$) before it turns over steeply at the inner scale \citep{Bas1994, Col1989, Sas2016, Sas2017, Mug2017}. Since our observations often sample scales in the vicinity of the dissipation scale, we use $\alpha = 3$.

The structure function discerned from observations is usually related to the spectrum of density fluctuations using expressions that are only valid in the limits where the observing baseline is either much larger than the dissipation scale ($s \gg l_{i}$) or much smaller than it ($s \ll l_{i}$) \citep{Col1987, Arm2000, Bas1994, Pra2011}. We use the general structure function (GSF \citep{Ing2014}) instead, which is valid in these limits, as well as in the intervening regime $s \approx l_{i}$:
\begin{equation}
\begin{split}
D_{\rm th}(s) & = C \times l_i(R)^{\alpha-2} \\
	   & {\times \bigg\{ { _1F_1} {\bigg[ - {{\alpha-2} \over 2},~1,~ - \bigg( {s \over l_i(R)} \bigg)^2 \bigg]} -1 \bigg\}} \, ,
\end{split}	    
\label{the_structfun}
\end{equation}

where
\begin{equation}\label{eq:gsf1}
C = {{8 \pi^2 r_e^2 \lambda^2 \Delta L} \over {\rho~2^{\alpha-2}(\alpha-2)}} {\Gamma \big( 1 - {{\alpha-2} \over 2} \big)} 
	    {{C_N^2 (R)} \over {(1 - f_p^2 (R) / f^2)}} \, ,
\end{equation}

\begin{figure}[!ht]
\centerline{\includegraphics[width=9.5cm]{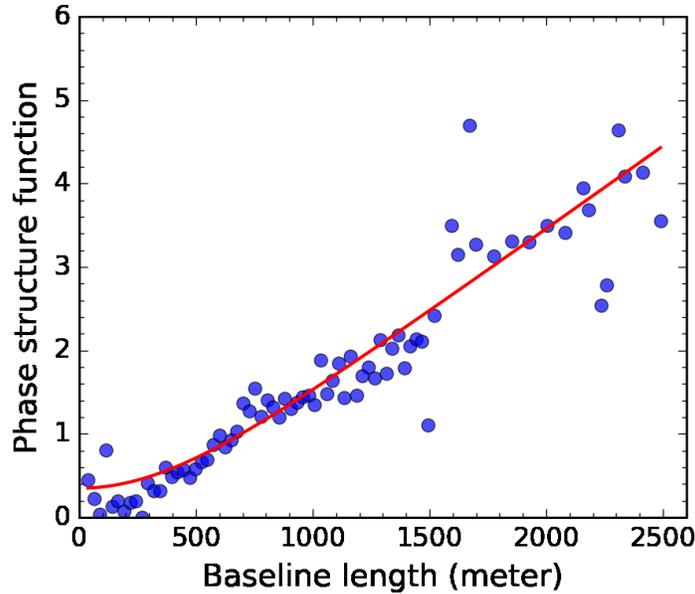}}
\caption{The circles indicate $D_{\rm obs}(s)/D_{\rm obs}(s_r)$ (Eq~\ref{obs_structfun}) for the Crab Nebula observed on 18 June 2016, the radio image for which is shown in Figure \ref{fig:aips_image}. The solid line represents $D_{\rm th}(s)/D_{\rm th}(s_r)$ (Eq~ \ref{the_structfun}) for $\alpha=3$ and the dissipation scale $l_i= 560$ m. The average rms error between the model fit (solid line) and the visibility data (circles) is 3.67\%}
\label{fig:fit}
\end{figure}

${ _1F_1}$ is the confluent hypergeometric function, $r_e$ is the classical electron radius, 
$\lambda$ is the observing wavelength, $R$ is the heliocentric distance, $\Delta L$ is the thickness of the scattering medium, $\rho$ is the axial ratio of the observed image, 
$f_p$ and f are the plasma and observing frequencies respectively. We choose a suitable reference baseline $s_r$ to define a function $D_{\rm th}(s)/D_{\rm th}(s_r)$. For a given value of the inner scale $l_i$, this is a function only of the baseline $s$.
The value of $l_{i}$ that minimizes the least squared difference between $D_{\rm th}(s)/D_{\rm th}(s_r)$ (Eq~\ref{the_structfun}) and $D_{\rm obs}(s)/D_{\rm obs}(s_r)$ (Eq~\ref{obs_structfun}) is regarded as our estimate for the dissipation scale. A successful fit for the observation of Figure~\ref{fig:aips_image} is shown in Figure \ref{fig:fit}. Results for all our observations are listed in Table \ref{tab:one}. Our estimates for $l_{i}$ in each case are listed in column (5) of Table \ref{tab:one}. The root mean square percentage difference between $D_{\rm th}(s)/D_{\rm th}(s_r)$ and $D_{\rm obs}(s)/D_{\rm obs}(s_r)$ for the estimated $l_{i}$ is listed in column (6). A small value for this number is a measure of the reliability of the estimate. 
The red circles in figure~\ref{fig:is} depict the estimated dissipation scale (column 7, table~\ref{tab:one}) as a function of (projected) heliocentric distance and the error bars on the red circles (which are typically smaller than the size of the plotting symbol) correspond to the fitting errors listed in column 8 of table~\ref{tab:one}.

Somewhat similar attempts have been made earlier - examples include \citet{Col1989} who used structure functions derived from radar observations, and \citet{Ana1994}, who have used structure functions derived from imaging observations such as the ones used in the present work. However, they estimate the dissipation scale ($l_i$) by looking for a break between the predictions of the asymptotic $s \ll l_{i}$ and $s \gg l_{i}$ branches of the structure function. Our method, where we fit the GSF (Eq~\ref{the_structfun}, which includes the asymptotic $s \ll l_{i}$ and $s \gg l_{i}$ branches together with the intervening $s \approx l_{i}$ regime) to the observed visibility data with $l_{i}$ as a fitting parameter, is considerably more accurate. We note that the baseline lengths used in our analysis (from 0.68 -- 36 km for VLA and 80 -- 2600 meters for GRAPH) are often comparable to the dissipation scale (column-5 of Table \ref{tab:one}), necessitating the use of the GSF.

\subsection{Comparing with dissipation scale models}\label{sec:sas2}

Most theoretical treatments of solar wind turbulence (especially those that attempt to address the kinetic regime) conclude that the turbulent spectrum first steepens at what is commonly referred to as the ``proton scale''. While some interpret the proton scale as the proton inertial length, some interpret it as the proton gyroradius (e.g., \citet{Bol2015}), while some conclude that the dissipation occurs at a range of scales \citep{Tol2015}. The proton inertial length model (\citet{Col1989, Yam1998, Bru2014}) envisages protons damping resonantly on $\rm Alf\acute{v}en$ waves, yielding the following expression for the dissipation scale:
\begin{equation}
 l_{i}(R) = d_i \equiv V_{A}/\Omega_{p} = 228~N_e(R)^{-1/2} \,\, {\rm km}\, \, ,
 \label{eq:inertiallength} 
\end{equation}
where $V_{A}$ is the $\rm Alf\acute{v}en$ speed, $\Omega_{p}$ is the proton cyclotron frequency and $N_{e}(R)$ is the electron number density in ${\rm cm}^{-3}$ at a heliocentric distance $R$. The solid line in figure~\ref{fig:is} depicts the prediction of the proton inertial length model with the electron density given by the model of Leblanc \citep{Leb1998}. The proton gyroradius is given by 
\begin{equation}
l_i (R) = 102 \mu^{1/2} {T_{i}(R)}^{1/2} {B(R)}^{-1} ~\rm cm,
\label{eq:protongyroradius}
\end{equation}
where $\mu$ is the ratio of mass of 
ion to mass of proton (taken to be 1 for our purposes), $T_i$ is the proton temperature in eV and B is the magnetic field in G. We use two models for the heliospheric magnetic field: one for a nominal Parker spiral magnetic field in the ecliptic e.g., \citep{Wil95},

\begin{equation}\label{eq:parker}
B(R)= 3.4 \times 10^{-5} R^{-2} (1+R^2)^{1/2} ~ \rm Gauss, \, ,
\end{equation}
where $R$ is the heliocentric distance in units of AU (1 AU = $215 R_{\odot}$),
and one using extrapolations using Helios observations \citep{Ven2018}, 

\begin{equation}
B (R) = 1.089 (0.0131\times SSN + 4.29)\times R^{-1.66} ~\rm nT \, ,
\label{eq:Bothmer}
\end{equation}
where $SSN$ refers to the sunspot number on the day of observation. The dissipation scale as a function of heliocentric distance using the proton gyroradius prescription of Eqs~(\ref{eq:protongyroradius}) and (\ref{eq:parker}) with a proton temperature of $10^{5}$ K is depicted by the red dashed line in figure~\ref{fig:is} while that using the prescription of Eqs~(\ref{eq:protongyroradius}) and (\ref{eq:Bothmer}) with a proton temperature of $10^{6}$ K is depicted by the blue dash-dotted line in figure~\ref{fig:is}. Barring the first point (at 2.2 $R_{\odot}$), our estimates for the dissipation scale are more consistent with a proton gyroradius interpretation than with the one that invokes the proton inertial length. The estimated value for the dissipation scale for the VLA observation of 1430-155 at 2.2 $R_{\odot}$ is substantially larger than those for the GRAPH observations, which were taken at larger heliocentric distances (column 7, table~\ref{tab:one}). There are three points to be noted in this connection - first, the sunspot number corresponding to the day on which the VLA observation was made was larger than those on which the GRAPH observations were made (column 5, table~\ref{tab:one}). Secondly, the source 1430-155 (triangle in figure~\ref{fig:schematic}) was located much closer to the solar south pole than the Crab nebula (circles in figure~\ref{fig:schematic}), suggesting that radiation from 1430-155 was predominantly sampling turbulence in the fast solar wind, while the Crab nebula observations were sampling (mostly) slow solar wind turbulence.
Finally, the short baseline (baselines ranging from a few 100 m to a few km) coverage for the VLA A array, with which the observations for 1430-155 were made, is much sparser than that for the GRAPH.
The relative lack of data points at short spacings tends to bias the model fit procedure (detailed in \S~3.1) towards higher values for $l_i$.

\section{Summary and Conclusions}

\begin{figure}[!ht]
\centerline{\includegraphics[width=9.5cm]{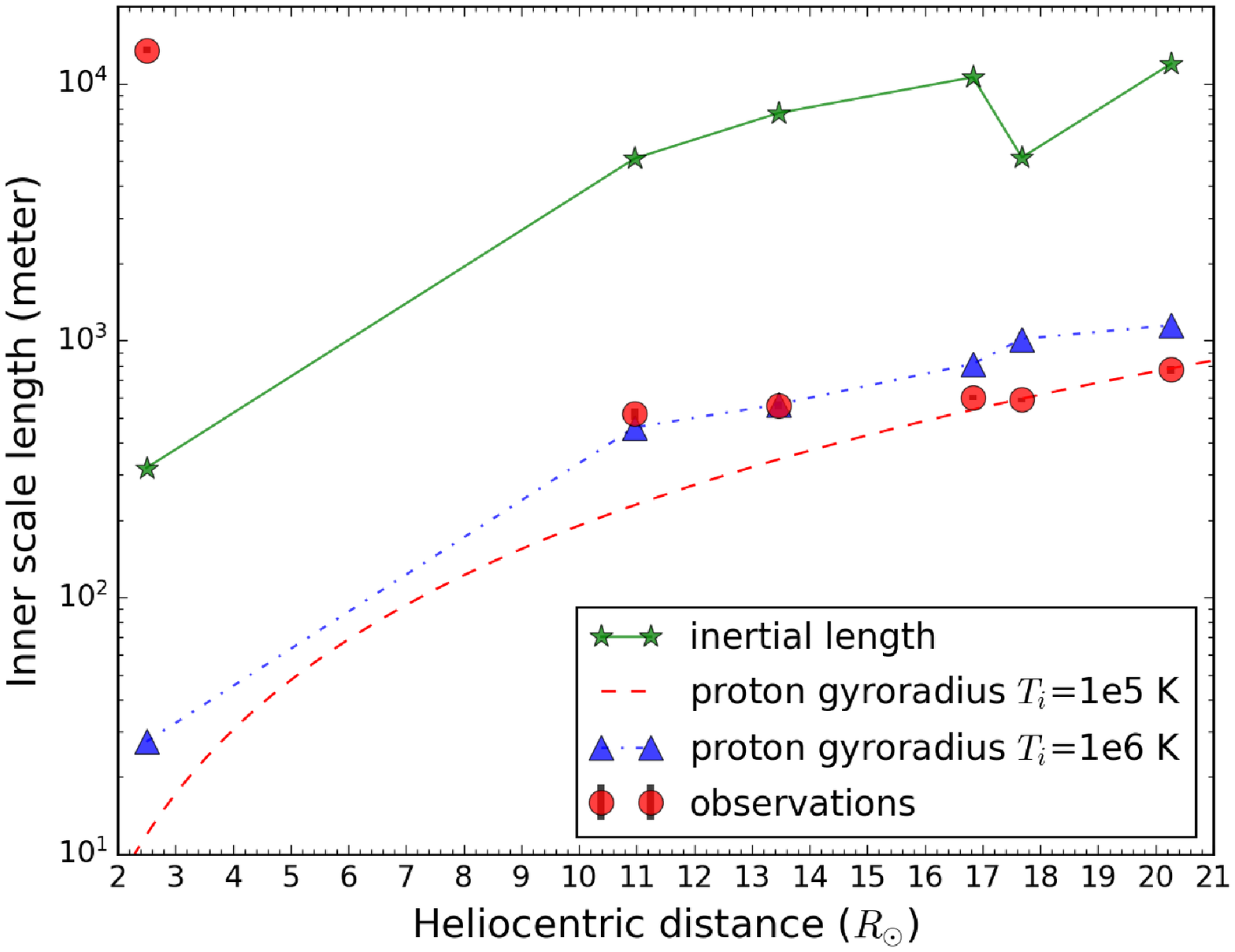}}
\caption{The dissipation length ($l_i$) of density turbulence 
in the solar wind (in meters) as a function of projected heliocentric distance in units of $R_{\odot}$. The red circles denote the dissipation scale lengths derived from the observations.
The solid line indicates the prediction of the proton inertial length model. The dashed line indicates the proton gyroradius computed using a temperature of $10^{5}$ K and the Parker spiral magnetic field model \citep{Wil95}, while the dash-dotted line indicates the proton gyroradius using a temperature of $10^{6}$ K and the magnetic field model of Venzmer \& Bothmer \citep{Ven2018}.
}
\label{fig:is}
\end{figure}

We have obtained reliable observationally based estimates of the dissipation scale of the turbulent density fluctuations in the solar wind for projected distances ranging from 2.5 -- 20.27 $R_{\odot}$ from the Sun center. The dissipation scale ranges from 
13500 to 520 m in this (projected) heliocentric distance range (Figure \ref{fig:is}). Our method uses observations of scatter-broadened sources observed against the turbulent solar wind. It does not invoke a specific model for the dissipation scale, and only assumes that the density turbulence spectrum comprises a power law in $k$-space together with an exponential turnover at the dissipation scale. However, we note that our estimates for $l_i$ for heliocentric distances ranging from 10--20 $R_{\odot}$ are generally consistent with the proton gyroradius. The value of the dissipation scale estimated at 2.2 $R_{\odot}$ departs significantly from this trend, and deserves special mention. This is especially important since it can be used to determine turbulent viscosity in a fluid model and (consequently) the extent of turbulent dissipation (i.e., heating) in the inner solar wind. As noted in \S~\ref{sec:sas1}, our dissipation scale estimate at 2.2 $R_{\odot}$ is more reliable than previous ones at similar heliocentric distances. As mentioned in \S~\ref{sec:sas2}, we also believe that the substantial discrepancy between the dissipation scale estimate at 2.2 $R_{\odot}$ and those at larger heliocentric distances has a physical origin. However, the relative lack of short baseline spacings in the observations contributing to the estimate at 2.2 $R_{\odot}$ might contribute to some observational bias. This issue can be effectively addressed by future observations of similar sources using radio interferometric arrays (or combinations thereof) which combine dense short baseline coverage with that at longer baselines. An instance of such observations (used to observe the solar corona) are those of \citet{Mer2006, Mer2015}.

Recent simulation studies suggest that the turbulence spectrum does not show a single dissipation scale as we assume; it first steepens at proton scales, and subsequently (i.e., at higher wavenumbers) at electron scales \citep{Ale2012, Mak2015}. In this context, our results for the dissipation scale can be interpreted as the first instance of steepening in the solar wind density turbulence spectrum. The physical nature of the density fluctuations is not yet clear. At large scales, where the MHD approximation is applicable, candidates include the fast \citep{Kol2018} and slow \citep{Bow2018} magnetoacoustic modes. At kinetic scales, (which is relevant to the dissipation scale lengths estimated in this work) a popular candidate in the solar wind context is the kinetic $\rm Alf\acute{v}en$ wave (KAW; e.g., \citet{Che2013}). There are also recent claims about direct observations of wave-particle coupling involving KAWs near the Earth's (dayside) magnetopause \citep{Ger2017}.

The Parker Solar Probe \citep{Fox2016} is expected to make detailed in-situ measurements of the solar wind at these distances, and our results can be expected to be a useful guide for interpreting them. Our results are expected to be useful in understanding dissipation mechanisms in MHD turbulence, and in pinning down the origin of extended solar wind acceleration. The dissipation scale is also an important determinant of the extent to which scattering due to refractive index inhomogeneities arising from density turbulence broadens sources - for sources located far behind the scattering screen of the solar wind, as well as for sources that are embedded in the solar corona \citep{Ing2014}. In the latter context, our estimates of the dissipation scale can be relevant in interpreting the scatter-broadening of type III burst sources observed with the LOFAR \citep{Kon2017} and other  longer interferometer baselines elsewhere \citep{Mug2016, Mug2018}.
\begin{table}
\centering
\vspace*{5px}
\begin{tabular}{|c|c|c|c|c|c|c|c|c| }
        \hline \hline

 S.&   Date & Source & R &  & Phase of the & $l_i$ & RMSE \\
 No. &   & Name   & ($R_{\odot}$) & SSN  & solar cycle&  (meter) & (\%)  \\
 (1) & (2) & (3) & (4) & (5) & (6) & (7) & (8) \\
    \hline
   1 &  1988 Nov 02 & 1430-155  & 02.5    & 143 & Ascending      & 13500   & 2.7  \\
   2 &  2016 Jun 18 & Crab      & 13.46   & 47  & Descending     & 560 & 3.67 \\
   3 &  2016 Jun 19 & Crab      & 16.83   & 50  & Descending     & 600 & 2.34 \\\
   4 &  2016 Jun 20 & Crab      & 20.27   & 36  & Descending     & 770 & 3.53 \\
   5 &  2017 Jun 10 & Crab      & 17.68   & 0   & Descending     & 590 & 2.0  \\
   6 &  2017 Jun 12 & Crab      & 10.97   & 0   & Descending     & 520 & 5.1 \\
   
       \hline \hline

\end{tabular}
\caption{Column 2 lists the date on which the scatter-broadened source was observed. Column 3 lists the source name and column 4 shows the projected heliocentric distance in units of the solar radius. Column 5 and 6 list the total sunspot number and phase of the solar cycle on that day. Column 7 lists the estimated dissipation scale length in meters while column 8 lists the percentage root mean square error (RMSE) associated with this estimate. 2017 Jun 10 and 12 were (Sun)spotless days, hence the zeroes corresponding to those days in column 5.}
\label{tab:one}
\end{table}

\acknowledgments
KSR acknowledges the financial support from the Science $\&$ Engineering Research Board (SERB), Department of Science $\&$ Technology, India 
(PDF/2015/000393). KSR acknowledges N.G. Kantharia and Eric Greisen
for valuable suggestions related to AIPS and M S Santhanam for valuable data analysis advice.
PS acknowledges support from ISRO RESPOND program.
The National Radio Astronomy Observatory is a facility of the National Science Foundation operated under cooperative agreement by Associated Universities, Inc.
We thank the staff of the Gauribidanur observatory for their help with the observations and maintenance of the antenna and receiver systems there. We thank the anonymous referee for an insightful report that has helped us improve the paper substantially.




\bibliographystyle{aasjournal}
\bibliography{ms}



\end{document}